\documentclass[%
 reprint,
altaffilletter,
 amsmath,amssymb,
 aps,
prstab,
longbibliography,
]{revtex4-1}

\usepackage{mathtools}
\usepackage{graphicx}
\usepackage{dcolumn}  % Align table columns on decimal point
\usepackage{bm}       % bold math

\usepackage{siunitx}
\usepackage{textgreek}

\usepackage{color}

\newcommand{\m}[1]{\SI{#1}{\meter}}
\newcommand{\mm}[1]{\SI{#1}{\milli\meter}}

\begin{document}

\title{Measurements of wake-induced electron beam deflection\\ in a dechirper at the Linac Coherent Light Source
}

\author{Johann Zemella}
 \altaffiliation[email: johann.zemella@desy.de\\]{Deutsches Elektron Synchrotron, 22603 Hamburg, Germany}%Lines break automatically or can be 
 \author{Karl Bane}
\author{Alan Fisher}
\author{Marc Guetg}
\author{Zhirong Huang}
\author{Richard Iverson}
\author{Patrick Krejcik} 
\author{Alberto Lutman}
\author{Timothy Maxwell}
\author{Alexander Novokhatski}
\author{Gennady Stupakov}
\author{Zhen Zhang}
\altaffiliation[]{Deparment of Engineering Physics, Tsinghua University, Beijing 100084, China}
\affiliation{SLAC National Accelerator Laboratory, Menlo Park, CA 94025, USA}
\author{Mark Harrison}
\author{Marcos Ruelas}
\affiliation{RadiaBeam Systems, Santa Monica, CA 90404, USA}

\begin{abstract}
The RadiaBeam/SLAC dechirper, a structure consisting of pairs of flat, metallic, corrugated plates,
%a corrugated structure in flat geometry, 
has been installed just upstream of the undulators
in the Linac Coherent Light Source (LCLS).
As a dechirper, with the beam passing between the plates on axis, longitudinal wakefields are induced that can remove unwanted energy chirp in the beam.
However, with the beam passing off axis, strong transverse wakes are also induced.
This mode of operation has already been used for the production of intense, multi-color photon beams using the Fresh-Slice technique, and is being used to develop a diagnostic for
attosecond bunch length measurements.
Here we measure, as function of offset, the strength of the transverse wakefields that are excited between the two plates, and also for the case
of the beam passing near to a single plate.
We compare with analytical formulas from the literature, and find good agreement. This report presents the first systematic measurements of the transverse wake strength in a dechirper, one that has been excited by a bunch with the short pulse duration and high energy found in an X-ray free electron laser. 

\end{abstract}

\maketitle

\section{Introduction}

The Linac Coherent Light Source (LCLS) \cite{EmmaP.2010} is a user facility at 
SLAC National Accelerator Laboratory %, CA, USA 
generating femto-second-long X-ray photon pulses.
The precise control of the electron beam parameters is essential for the optimal performance 
of the free electron laser (FEL) \cite{PhysRevSTAB.6.050702,PhysRevSTAB.9.050702}. 

In order to generate X-ray FEL pulses, high slice electron density, low slice emittance, and
low slice energy spread are needed \cite{FELBuch}.
At the LCLS the high current is achieved 
via a staged magnetic compression scheme. 
%An off-crest RF acceleration provides a longitudinal position to energy correlation within the bunch; this is followed by a chicane, which results in 
%compression of the bunch.
Normally, however, an uncorrected longitudinal position to energy correlation---an energy chirp---remains after the bunch has left the final compressor chicane.
An idea was developed of using a metallic structure with small corrugations for chirp control in X-ray FELs~\cite{dechirper}. When placed after the linac, such a  ``dechirper'', through wakefields induced by the beam, can passively remove unwanted energy chirp before the beam enters the undulator. One such device, the RadiaBeam/SLAC dechirper, was installed in the LCLS at the end of 2015. It consists of two modules, each comprising two, flat adjustable corrugated plates with the beam passing in between.

In addition to the longitudinal wake effect in the dechirper, the induced transverse wake effects are strong and cannot be ignored. 
If the beam moves through the device off axis, a strong transverse kick is induced that is largest at the tail of the bunch; but even if the beam passes through the structure on axis, strong induced quadrupole wakes can greatly increase the beam's projected emittance.
Nevertheless, recently there has been success in taking advantage of the strong transverse wake forces. Dechirper transverse wakefields were used to selectively control the electron bunch 
lasing slice at the LCLS, allowing for the production of intense, multi-color photon beams 
using the Fresh-Slice technique~\cite{Alberto}.
A similar idea is being investigated, one that has the beam on axis and uses the quad wake to select the lasing slice~\cite{mismatch}. 
Also, temporal diagnostic schemes based on the transverse wakes have been demonstrated at low energies  
\cite{PhysRevAccelBeams.19.021304}, and hold the promise of an attosecond electron bunch diagnostic 
operating at high energies~\cite{PhysRevSTAB.18.104402}. 

The measurement of the wakefields---both longitudinal and transverse---of a flat plate dechirper was performed before~\cite{PAL}; 
however, the beam energy and current were low: $\sim70$ MeV and $\sim40$~A, respectively.
%however, the bunch length was long, $\sim400$~$\mu$m (rms), and the beam energy was low,  $\sim70$ MeV.   
The same can be said for detailed measurements of the quad wake of a dechirper~\cite{quad_wake_meas}. In contrast, the measurements presented here have the energies ($\gtrsim6$~GeV) and currents (several~kA's) of a functioning X-ray FEL. Since the bunch is much shorter, we probe the impedance to much higher frequencies than has been done before. Note that some of the results presented here, as well as additional aspects of the measurements, can be also found in Ref.~\cite{dechirper_commissioning}.

In this report, we present measurements of the deflection of an electron bunch: (1)~as it passes off axis between two plates of the RadiaBeam/LCLS dechirper, as a function of offset, and (2)~as it passes near a single plate of the dechirper, as a function of beam offset. We will compare the measurements with analytical formulas derived in Ref.~\cite{PhysRevAccelBeams.19.084401}. Note that these formulas were confirmed to agree well with numerical simulations (for the typical LCLS bunch lengths) using the computer program Echo(2D)~\cite{PhysRevSTAB.18.104401}. 
In the present work we also report on measurements of the transverse kick experienced by a beam passing close to a single dechirper plate. These results are compared with wake formulas found in~\cite{SLAC-Pub-16881}. %Finally, all measurement results will be compared with numerical results of the computer program NOVO [Ref]. 
%Note that in the two-plate case, from the symmetry of the data we can verify the axis location; in the single-plate case, however, we have no such symmetry. Here we fit to an unknown shift in beam offset. To give us confidence in the single-plate results, we, in addition, simultaneously measure the longitudinal wake effect and again fit to an unknown shift; and in the end we compare the two shifts.

This report is organized as follows.
In Sec.~\ref{sec:theo} we present the analytical wake formulas that are used to compare to the measurements. 
In Sec.~\ref{sec:expe} the experimental setup is described. 
In Sec.~\ref{sec:two} we present two-plate measurement results and compare to theory, and in Sec.~\ref{sec:one} we do the same for the single plate measurements. 
%The ends of the jaws are independently adjustable; we begin this section with the study of how to align the jaws and the effect of jaw tilt. 
The conclusions come in Sec.~\ref{sec:conc}. 
In Appendix~A we derive the average kick for a uniform bunch from the wakefield formulas, and also present longitudinal results needed for the single plate comparisons.

\section{Theory of wakefield-induced electron deflection}
\label{sec:theo}
%\input{theo_deflectionb.tex}

%The transverse motion of a particle at longitudinal position $s$ in a bunch can be parameterized by its transverse position $(x,y)$ 
%and angle $(p_x,p_y)$.
The concept of wakefield is useful for the study of beam dynamics of high energy, charged-particle beams in an accelerator.
Leading charged particles generate an electromagnetic field (wakefield) that interacts with the boundary and then acts on trailing particles.
For a high energy beam passing through a device, like a dechirper, the effect can be described by an impulse kick on each particle.
For a line charge beam moving at offset $x$ parallel to the $z$ axis of such a structure,  
the kick induced at longitudinal position $s$ (within the bunch) by a transverse wakefield is described by the convolution 
%(see e.g. \cite{LCC-0116})
\begin{align}
 V_{x}(s,x) &=QL \int_{-\infty}^{s}  
 w_{x}(s-s',x) \lambda(s')\,ds'\ ,
 \label{eq:tKick}
\end{align}
with $Q$ the bunch charge, $L$ the length of structure, $w_{x}(s,x)$ the transverse, point charge wakefield (in units of V/pC per unit length), and 
$\lambda(s)$ the longitudinal bunch distribution (normalized to unit area). 
Here the head of the bunch is located in the negative $s$ direction, with $s_{head} < s_{tail} $.

\subsection{Wakes for a pencil beam in flat geometry}
\label{sec:SOA}

Our measurements concern ``pencil beams," {\it i.e.} beams with a transverse extent that is small compared to the aperture or distance to the nearest wall.
Consider a vacuum chamber object that has flat geometry, oriented as the two plates of the horizontal dechirper used in the measurements of this report (the plates are parallel to the $x=0$ plane, which defines the symmetry axis). For small offsets from the axis, the transverse point charge
wake can be written in the form~\cite{PhysRevSTAB.18.034401}
\begin{align}
w_{x}(s)=x_{0}\, \bar w_{d}(s) + x\, \bar w_{q}(s)\ ,\quad
w_y(s)=\bar w_q(s)(y_0-y)\ ,
\end{align}
with ($x_0$, $y_0$), ($x$,$y$), respectively the offset of the leading and trailing particle; with $\bar w_d(s)$, $\bar w_q(s)$, respectively the dipole and quadrupole wake functions (the over-bars indicate that these functions are normalized to offset and thus have different units than $w_{x,y}$).  
For the case of two particles still near each other transversally but away from the axis, with the leading particle at offset $x_0$ and the trailing one at $x$ (with $|x-x_0|$ small compared to the  the distance to the nearest wall) , the wakes are of the form~\cite{PhysRevAccelBeams.19.084401}
\begin{align}
&&w_{x}(s)= w_{d}(s,x_0) + \bar w_{q}(s,x_0)(x-x_0)\ ,\nonumber\\
&&w_y(s)= \bar w_q(s,x_0)(y_0-y)\ .\label{wx_dipolequad_eq}
\end{align}
In this case, both the dipole and quad wake components depend non-linearly on the offset $x_0$. For the case of a beam near a single plate at offset $b$, the form of the wakes is the same but with the parameter $x_0$ replaced by $b$.

The transverse wakes at the origin in $s$ (at $s=0^+$) start with a linear slope. In a two-plate dechirper of gap $2a$, for a beam near the axis, the slopes at the origin of the dipole and quad wake functions are given by~(see {\it e.g.} \cite{PhysRevSTAB.18.034401}) 
\begin{align}
\bar w'_{d} = \bar w'_{q}= \frac{\pi^3}{128} \frac{Z_0 c}{a^4}\ ,
\end{align}
with  $Z_0=377$~$\Omega$ and $c$ the speed of light.
For a pencil beam passing by a single dechirper plate, the slopes of the transverse wakes at $s=0^+$ are given as~\cite{SLAC-Pub-16881}
\begin{align}
w'_{d} = \frac{Z_0 c}{4\pi b^3}\ ,\quad\quad \bar w'_{q}=  \frac{3Z_0 c}{8\pi b^4}\ .
\end{align}

\subsection{Analytical formulas for short bunches }
\label{sec:IgorTheo}

\begin{figure}[htb]
\centering
\includegraphics[width=0.3\textwidth, trim=0mm 0mm 0mm 0mm, clip]{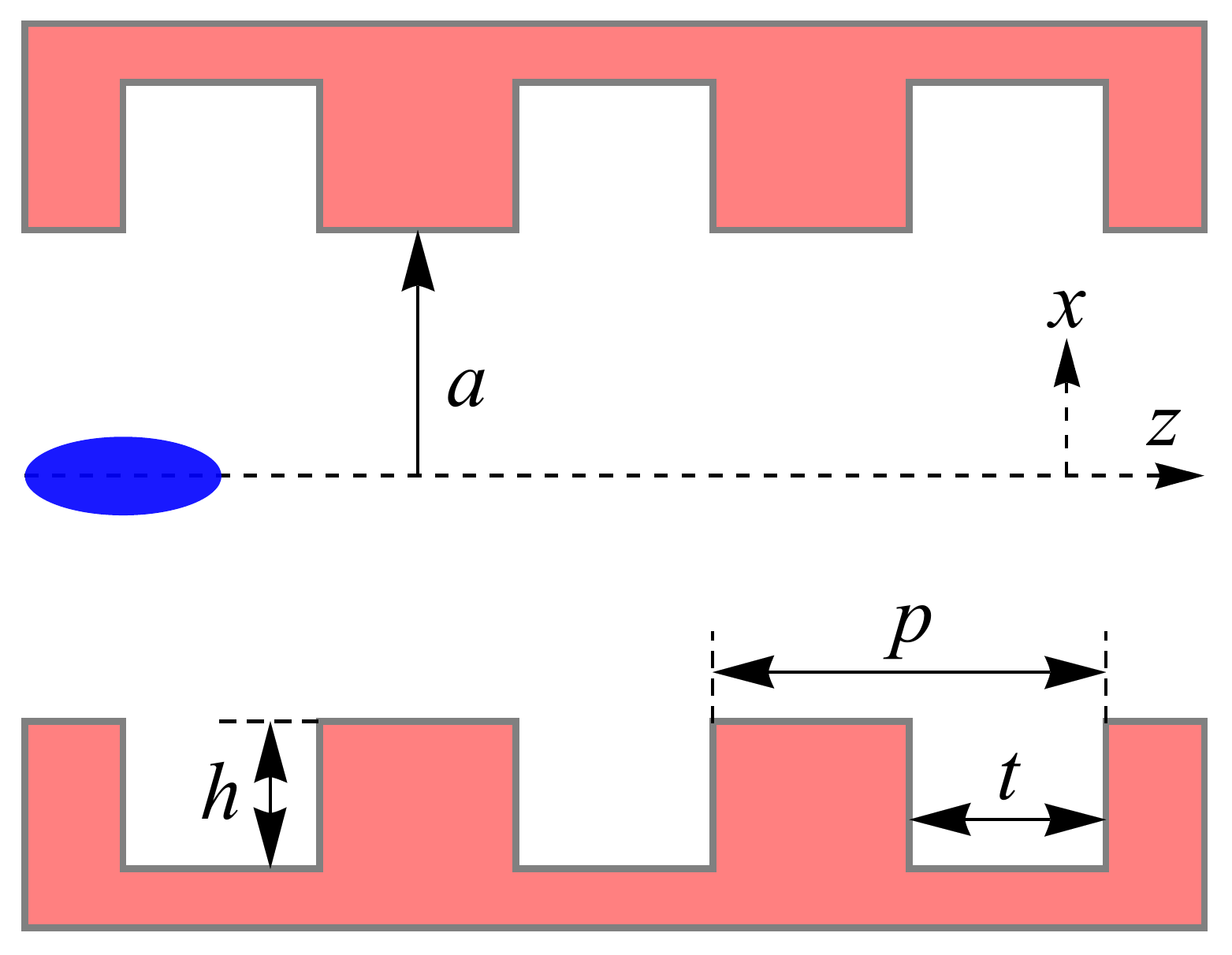}
\caption{Three corrugations of a horizontal dechirper. A rectangular coordinate system is centered on the symmetry axis of the chamber. The blue ellipse represents an electron beam propagating along the $z$ axis.}
\label{fig:1}
\end{figure}

The dechirper geometry is sketched in Fig.~\ref{fig:1}. The parameters are half gap $a$ and corrugation parameters: period $p$, longitudinal gap $t$, and depth $h$. Ideally, the corrugation parameters of a dechirper are small ($h$, $p\ll a$), but with $h\gtrsim p$~\cite{dechirper}.
The equations described below apply for the type of parameters of the RadiaBeam/SLAC dechirper and the short bunches at the end of the LCLS. They are the equations for a horizontal dechirper, {\it i.e.} one with plates parallel to the $x=0$ plane, 
like the one shown in the figure.

The transverse point charge wake, both for the cases of the beam at offset $x$ between two plates and the beam at distance $b$ from a single plate, is given in the form~\cite{PhysRevAccelBeams.19.084401}
\begin{equation}
w_x(s)=\left(\frac{Z_0c}{4\pi}\right)As_{0x}\left[1-\left(1+\sqrt{s/s_{0x}}\right)e^{-\sqrt{s/s_{0x}}}\right]\ ,\label{wy_eq}
\end{equation}
with $A$ and $s_{0x}$ constants that depend on the aperture, the beam offset, and the corrugation parameters. Here the leading and trailing particles are taken to have identical transverse offsets; thus $w_x$ equals the first, dipole term of Eq.~\ref{wx_dipolequad_eq}. For two plates separated by distance $2a$, with the beam offset from the axis by $x$, the constants are (subscript $d$ stands for ``double'' plates)
\begin{eqnarray}
A_d&=& \frac{\pi^3}{4a^3}\sec^2\left(\frac{\pi x}{2a}\right)\tan\left(\frac{\pi x}{2a}\right)\ ,\nonumber\\
s_{0xd}&=&4s_{0r}\left[\frac{3}{2}+\frac{\pi x}{a}\csc\left(\frac{\pi x}{a}\right)-\frac{\pi x}{2a}\cot\left(\frac{\pi x}{a}\right)\right]^{-2}\, ;\label{Ad_eq}
\end{eqnarray}
the scale factor
\begin{equation}
s_{0r}=\frac{a^2t}{2\pi \alpha^2(t/p)p^2}\ , 
\end{equation}
where $\alpha(\zeta)=1-0.465\sqrt{\zeta}-0.070\zeta$.

In the case of a beam passing by a single plate at distance $b$, the constants are given by (subscript $s$ stands for ``single'' plate)~\cite{SLAC-Pub-16881}
\begin{equation}
A_s=\frac{2}{b^3}\ ,\quad\quad s_{0xs}=\frac{8b^2t}{9\pi \alpha^2 p^2}\ .
\end{equation}

When the beam passes the dechirper off axis, it receives a transverse wake kick, which translates to a change in offset at a downstream beam position monitor (BPM). In the experiments described below we record the change in offset at such a BPM, which actually measures the offset change averaged over the longitudinal bunch distribution. Since, in our case, there is only a drift region between dechirper and measuring BPM, the change in (averaged) offset is simply given as 
\begin{equation}
\Delta x_w=eQ\varkappa_xLL_{BPM} /E\ , \label{deltax_eq}
\end{equation}
with $e$ the charge of an electron, $L$ the length of the dechirper plate, $L_{BPM}$ the distance from dechirper to the measuring BPM, $\varkappa_x$ the kick factor (the average of the transverse bunch wake), and $E$ the bunch energy. The bunch at the end of the LCLS is short and, to good approximation, has a uniform bunch distribution. The expression for $\varkappa_x$, for a short, uniform bunch of full length $\ell$, is derived in Appendix~A and given as
\begin{equation}
\varkappa_x=\left(\frac{Z_0c}{8\pi}\right)As_{0x}f_x\left(\frac{\ell}{s_{0x}}\right)\ ,\label{kappax_eq}
\end{equation}
with
\begin{equation}
f_x(\zeta)=1-\frac{12}{\zeta}+\frac{120}{\zeta^2}-8e^{-\sqrt{\zeta}}\left(\frac{1}{\zeta^{1/2}}+\frac{6}{\zeta}+\frac{15}{\zeta^{3/2}}+\frac{15}{\zeta^2}\right)\ .
\end{equation}

The parameters of the RadiaBeam/LCLS dechirper are given in 
Table~\ref{tab:paraCS}. Note that at the smallest values of $a$, the corrugation parameters are not small ($h$, $p\not\ll a$).

\begin{table}
 \centering
 \caption{Parameters of the RadiaBeam/SLAC dechirper.}
 \label{tab:paraCS}
 \begin{tabular}{lcc}
  Parameter & Value & Unit \\
  \hline
  Half gap $a$ & $0.5$--$12.5$      & \mm{} \\
Jaw overtravel  & 3      & \mm{} \\
  Corrugation properties:\\
  \hspace{4mm}Period $p$         & $0.5$           & \mm{} \\
  \hspace{4mm}Depth  $h$          & $0.5$           & \mm{} \\
  \hspace{4mm}Longitudinal gap  $t$        & $0.25$          & \mm{} \\
  Plate width $w$        & $12.$          & \mm{} \\
  Length of structure $L$ & $2$      & \m{}
 \end{tabular}
\end{table}

\section{Experimental setup}
\label{sec:expe}

\begin{figure*}
   \centering
   \includegraphics[width=1\textwidth]{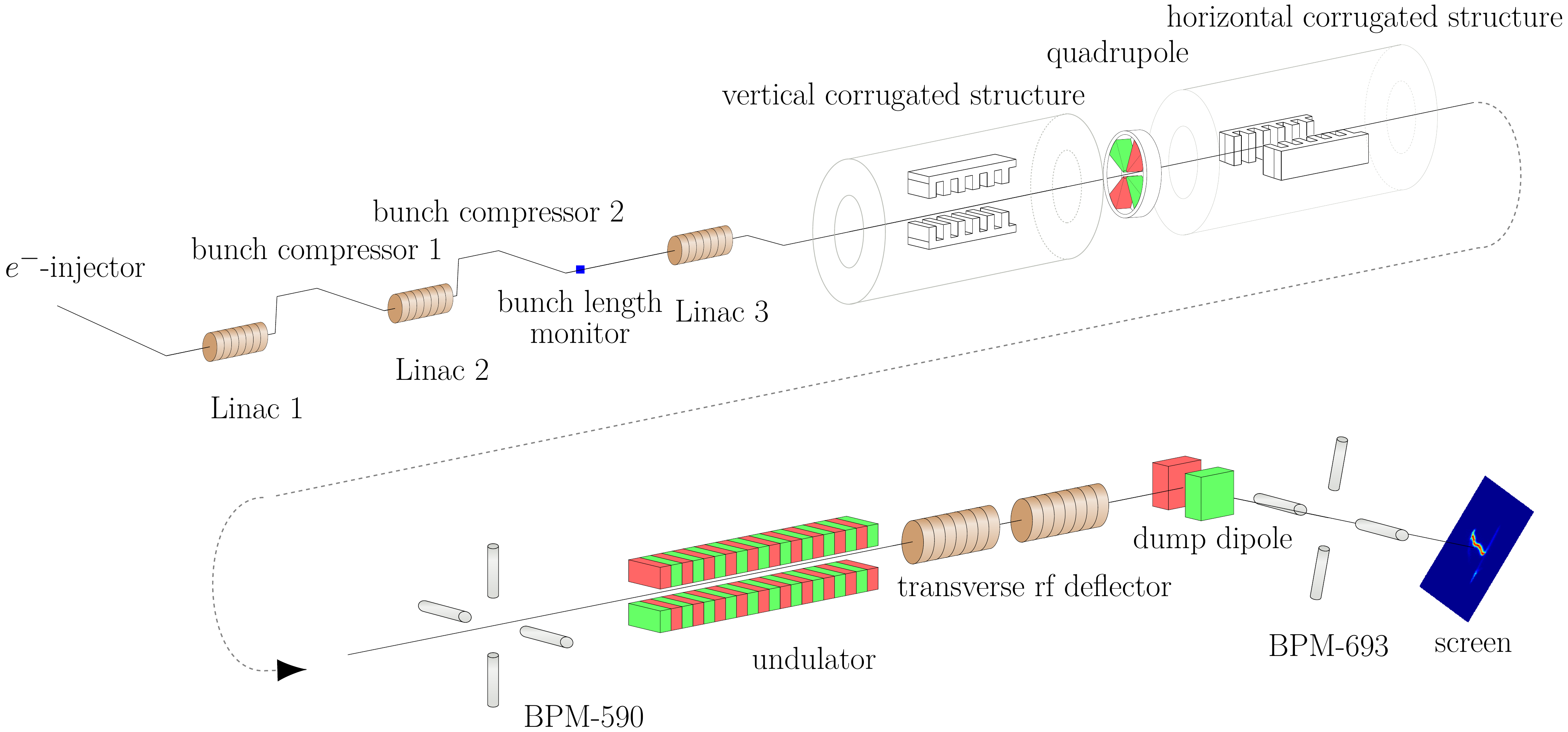}
   \caption{Sketch of the the LCLS beam line, focusing on the  RadiaBeam/SLAC dechirper and the diagnostics used in the measurements of this report. Note that the relative scale of distances is greatly distorted; {\it e.g.} in reality the linacs are much longer than the other objects in the figure. During the measurements, the first, vertical dechirper module is kept wide open and the second, horizontal one's effect on the beam is investigated.
  }

  \label{fig:expSetup}
\end{figure*}

In Fig.~\ref{fig:expSetup} we sketch the LCLS beam line, where we focus on the  RadiaBeam/SLAC dechirper and the diagnostics used in the measurements of this report.
At the right in the top row, after the injector and linacs, we show the dechirper region, including the vertical dechirper module, followed by a quadrupole, and then the horizontal dechirper module.
Note that the relative scale of distances in the sketch is greatly distorted; {\it e.g.} in reality the linacs are much longer than the other objects in the figure. 
The measurements described in this report have the vertical dechirper jaws opened wide, and use only the horizontal module, which kicks the beam horizontally (in $x$). 
%The diagnostics used in the measurements include: coherent edge radiation bunch length monitor at the end of bunch compressor 2 (BC2) [used in] bottom row of Fig.~\ref{fig:expSetup} shows selected accelerator components downstream of the dechirper that are involved in the measurements.
In our measurements, the first BPM past the dechirper, BPM 590, at a distance of $L_{BPM}=16.26$~m from the dechirper center, is used to measure the wake kick; note that there is only drift space between dechirper and BPM.
%During the deflection measurements the beam is dumped upstream of the undulator. 

During the two-plate measurements, described first below, the beam was very stable. 
After the transverse scans in the dechirper were taken,
the current profile of the beam was measured using the X-band
transverse RF-deflector (XTCAV)~\cite{PhysRevSTAB.14.120701} and its screen, both located downstream of the undulator. The bunch form was needed for
the theoretical calculations.

During the single plate measurements, described below, the shot-by-shot bunch length is measured using the coherent edge radiation generated in the last bend of the final bunch compressor, BC2 (located just before Linac~3; see Fig.~\ref{fig:expSetup})~\cite{edge_radiation}. To compare with these measurements, we use the formulas assuming a uniform bunch distribution. Supporting longitudinal measurements use BPM 693, located downstream in the beam dump region, to measure energy change (where dispersion $\eta_x=57.7$~cm). 

It is important to note that in the RadiaBeam/SLAC dechirper, the jaws are independently adjustable at both ends. Although, in setup, we normally endeavor to align them parallel to each other and to the beam motion, we expect to inevitably end up with a small amount of residual taper.

%\vspace{8mm}
\section{Two-plate measurements}
\label{sec:two}
%\input{meas_deflection.tex}

%two plate parallel

%\subsection{Two-plate measurements with parallel jaws}

\begin{figure}[!htb]
    \centering
    \includegraphics[draft=false, width=.48\textwidth]{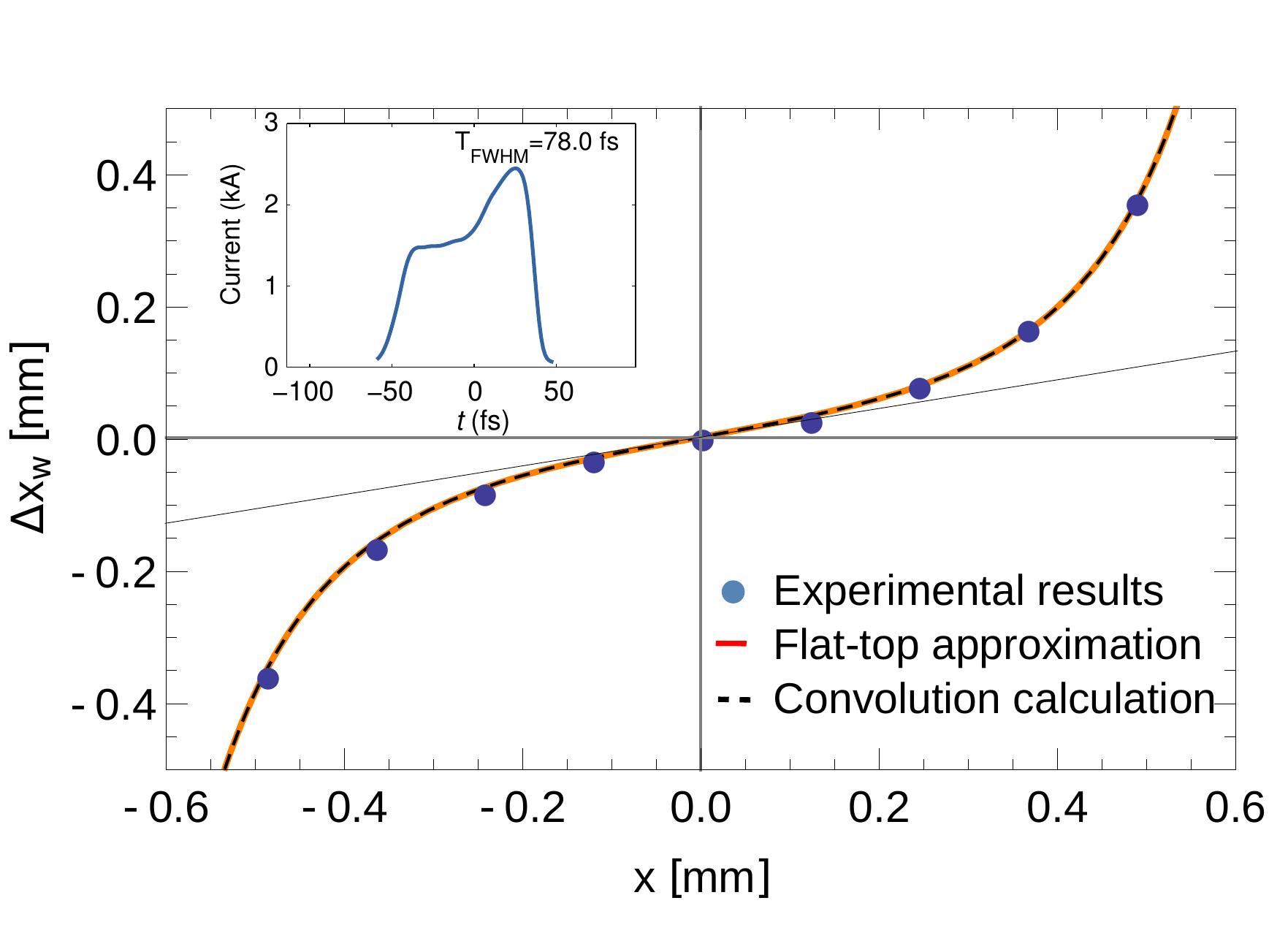}
    \caption{Deflection as function of beam offset from the axis in the horizontal  dechirper module, showing data (plotting symbols) and calculations (curves), for half gap $a=1$~mm. Note that the error bars in the data are small and not visible.  For the simulated curves, the gap parameter was reduced by 11\% to fit the experimental data. The tilted straight line gives the slope of the curves at the origin. The bunch shape, with head to the left, is given in the inset figure.
  %  \color{red}This plot needs the bunch shape as an inset, and a legend showing blue symbol ``Experimental results'', red line ``Flat-top approximation'', black dashes ``Convolution current profile''.\color{black}
    \label{two_plates_a1mm_fi}
  }
    \end{figure}

For the first measurement we set the half gap between the jaws to $a=1$~mm. Here charge $Q=152$~pC and energy $E=6.6$~GeV. We keep the beam fixed, scan the dechirper across the beam, and measure the offset at BPM 590, $\Delta x_w$. The results are given in Fig.~\ref{two_plates_a1mm_fi}, where the abscissa is the offset of the beam with respect to the dechirper axis, $x$ (a positive offset gives a positive kick). The figure shows the data (plotting symbols with error bars) and calculations (the curves).  Each data point represents the average of 40 shots, with error bars showing $+/-$ one standard deviation. (However, in this and some following plots, the error bars are so small as to not be visible.) The bunch shape (with head to the left) as obtained from the XTCAV screen, is given in the inset figure.

  %\vspace{6mm}

 The first thing that we note is that the data set is good, in that the nine data points trace out a smooth, nearly perfectly anti-symmetric curve about the origin, which is theoretically required in a two-plate dechirper that is symmetric in $x$. 
 %Also, the standard deviation of the data is small ($\le0.006$~mm), and the error bars are not visible in the figure. 
For the scale of the wake effect, note that a value of $\Delta x_w=0.4$~mm corresponds to an average kick of $V_x=160$~kV. 
    
    For the comparison calculations, we first numerically performed the wake convolution (Eq.~\ref{eq:tKick}) with the measured bunch shape and the wake function (Eqs.~\ref{wy_eq}, \ref{Ad_eq}), to obtain the voltage $V_x$; then the result is given by $\Delta x_w=eV_xL_{BPM}/E$. In a second calculation, we used the analytical solution assuming a uniform bunch distribution (Eqs.~\ref{deltax_eq}, \ref{kappax_eq}) to obtain the deflection  $\Delta x_w$.
In this case we need an effective full bunch length $\ell$. This parameter was obtained from the measured bunch shape as $\ell=2\sqrt{3}ct_{rms}$, with $t_{rms}$ the numerically obtained, rms width (in time) of the bunch distribution. This calculation yields an effective peak current of $I=1.76$~kA and full length of $\ell=25.9$~$\mu$m.

In Fig.~\ref{two_plates_a1mm_fi} we plot the two calculation results on the same plot as the data. However, to obtain a good fit to the data, we changed the half-aperture parameter $a$ in the calculations by $-11$\%. The straight slanted line in the figure gives the slope of the calculations at the origin.
We note that the analytical formula assuming a uniform bunch distribution (the orange curve) is almost identical to the more involved convolution calculation (the black, dashed line). The analytical formula also agrees well with the measured data, except for the $-11$\% scale factor. At the moment we do not understand the source of the discrepancy. If it is an error in the measurement, it may be due to the inaccuracy of knowing the jaw separation and/or an unknown tilt in the jaws.

\begin{figure}[!htb]
    \centering
    \includegraphics[draft=false, width=.48\textwidth]{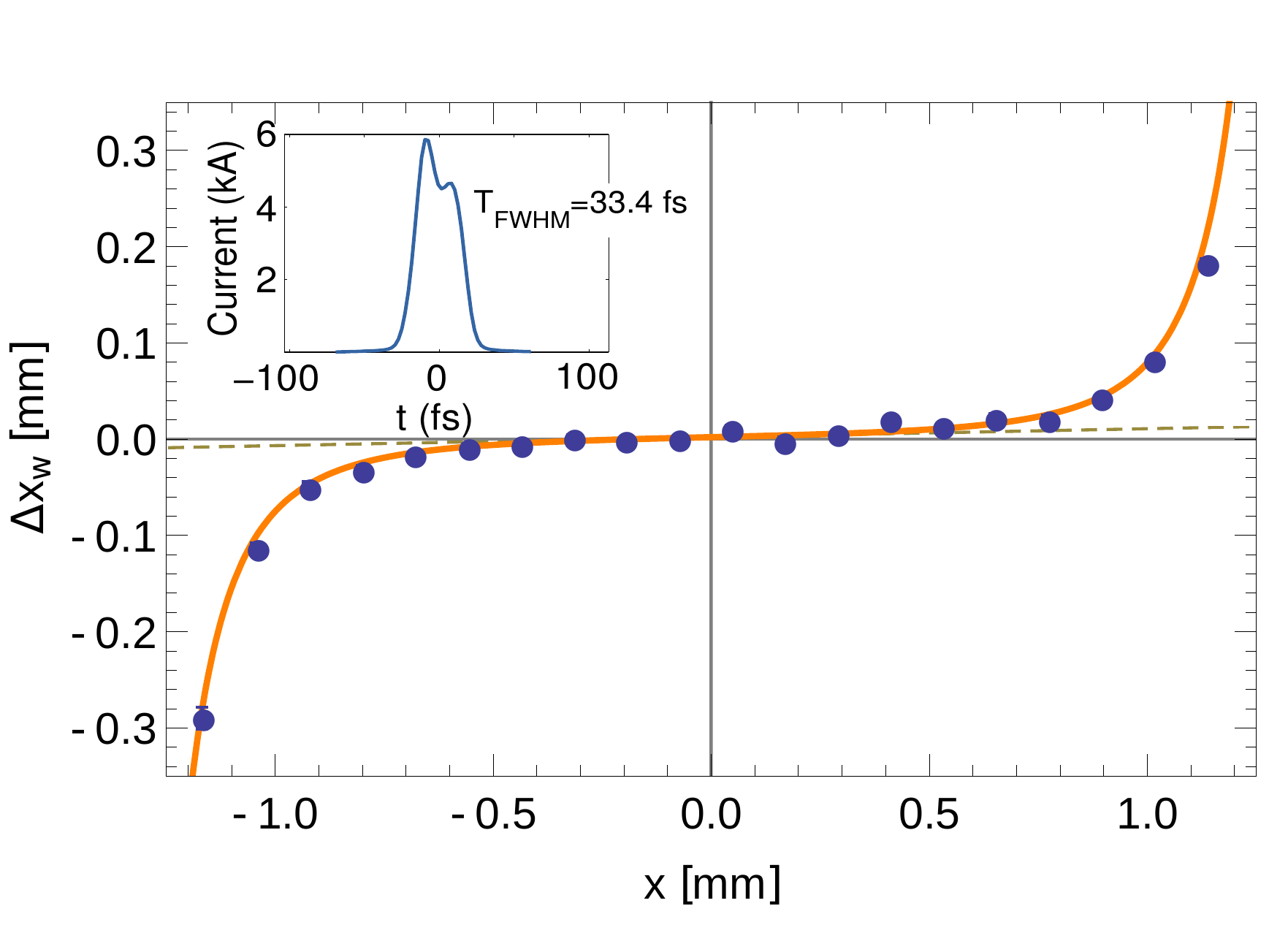}
    \caption{Deflection as function of beam offset from the axis in one dechirper module, showing data (plotting symbols) and calculations (flat-top approximation, the curve), for half gap $a=1.55$~mm.  Note that the error bars in the data are small and barely visible. The gap in the analytical calculation was reduced by 6\% to fit the experimental data. The straight line gives the slope of the curve at the origin. The bunch shape, with head to the left, is given in the inset figure.
   % \color{red}This plot needs the bunch shape as an inset.\color{black}
    \label{two_plates_a1p55mm_fi}
  }
    \end{figure}

We repeat the measurement, but now with half gap $a=1.55$~mm, charge $Q=187$~pC, and energy $E=13.3$~GeV. The data is shown in Fig.~\ref{two_plates_a1p55mm_fi} (the plotting symbols), with the error bars again small and barely visible. Note that the data points, at first, were shifted upward in the figure, by 40~$\mu$m, to yield zero kick on axis; and then, to make the data anti-symmetric, they were shifted to the right by 35~$\mu$m. Note that a value of $\Delta x_m=0.3$~mm here corresponds to a kick of $V_x=245$~kV.

   For the comparison calculation, we used the analytical solution assuming a uniform bunch distribution.
From the bunch shape measured by XTCAV (see inset) we find the effective peak current and full bunch length as before; the results are $I=4.33$~kA and $\ell=13.0$~$\mu$m.  To obtain a good fit to the data, we changed the half-aperture parameter $a$ in the calculations by $-6$\%.
The straight, dashed line gives the slope of the wake at the origin, which we see is much smaller than before. The wake effect over most of the measured range is quite small; it only begins to become significant for $|x|\gtrsim1$~mm. The agreement between theory and measurement is quite good.

%single plate

\section{Single jaw measurements}
\label{sec:one}

With the beam passing by a single jaw of the dechirper, its offset at BPM 590, $\Delta x_w$, was measured as function of distance of the beam from the jaw, $b$. One set of measurements used the north jaw of the horizontal dechirper module, the other used the south jaw. Careful setup of the beam allowed the electron bunch to be much closer to the jaw (than in the two-plate examples above), resulting in a stronger induced wakefield. The recording of $\sim1000$ measurements per scan step allowed for aggressive filtering with respect to incoming orbit and beam current (using the coherent edge radiation, bunch length monitor). For the north (south) jaw data presented below, we ended up keeping more than 25 (50) data points, except when the beam was  0.5~mm or closer to the jaw, when the number of good points dropped to $\sim 10$.

 The beam parameters were charge $Q=180$~pC, peak current $I=3.5$~kA, which implies that the full bunch length $\ell=14$~$\mu$m; and energy $E=13$~GeV. 
To clearly know the zero in kick, we included data points with the beam many millimeters from the jaw. Unlike in the two-plate measurements, in the single plate case we can't use symmetry to help judge the goodness of the data. 
A fit of the model, Eqs.~\ref{deltax_eq}, \ref{kappax_eq}, to the measurements was performed, where we allow for an overall shift in $b$, denoted by $\Delta b$.

At the same time as measuring the transverse kick, as a consistency check,
%via the offset of the beam at a downstream BPM, 
we measured the longitudinal wake effect using a downstream BPM in a dispersive region, BPM 693. This gives us the average energy loss of the beam to the wakefields. Fitting this data to the longitudinal formulas in the same way, and comparing the fitted offset with that of the transverse measurement gives us a better idea of the accuracy of the analytical models. From the measurements we obtain the energy loss as $\Delta U=x_m E/\eta_x$, with $x_m$ the measured offset and $\eta_x$ the dispersion at the BPM 693 (here $\eta_x=57.7$~cm). For the calculations, it is given by $\Delta U=eQL\varkappa$, with the loss factor $\varkappa$ derived in Appendix~A, in Eq.~\ref{kappa_eq}.

In Fig.~\ref{north_kick_fi}, the top plot, we present the north jaw transverse measurement results. Note that for the single plate measurements, $\Delta x_w>0$ indicates a kick toward the jaw. A value of $\Delta x_m=0.6$~mm corresponds to a kick of $V_x=480$~kV.
%We note that the error bars are again small. In addition, 
We see that the fit of the theory (the curve) to the data (the plotting symbols) is good. The fit gives an overall shift of $\Delta b=-161$~$\mu$m, and the data in the plot has been shifted by this amount.
 In Fig.~\ref{north_kick_fi}, the bottom plot, we present the north jaw energy measurements. We note that there is more scatter in the energy loss data than in the transverse kick data. We see that the fit of the theory to the data again is good. The fit gives an overall shift of $\Delta b=-138$~$\mu$m (and, like before, the data in the plot has been shifted by this amount). We see that, in both cases, the theory agrees well with the measurements, and the fitted shifts also are in reasonable agreement with each other.

 \begin{figure}[!htb]
    \centering
    \includegraphics[draft=false, width=.45\textwidth]{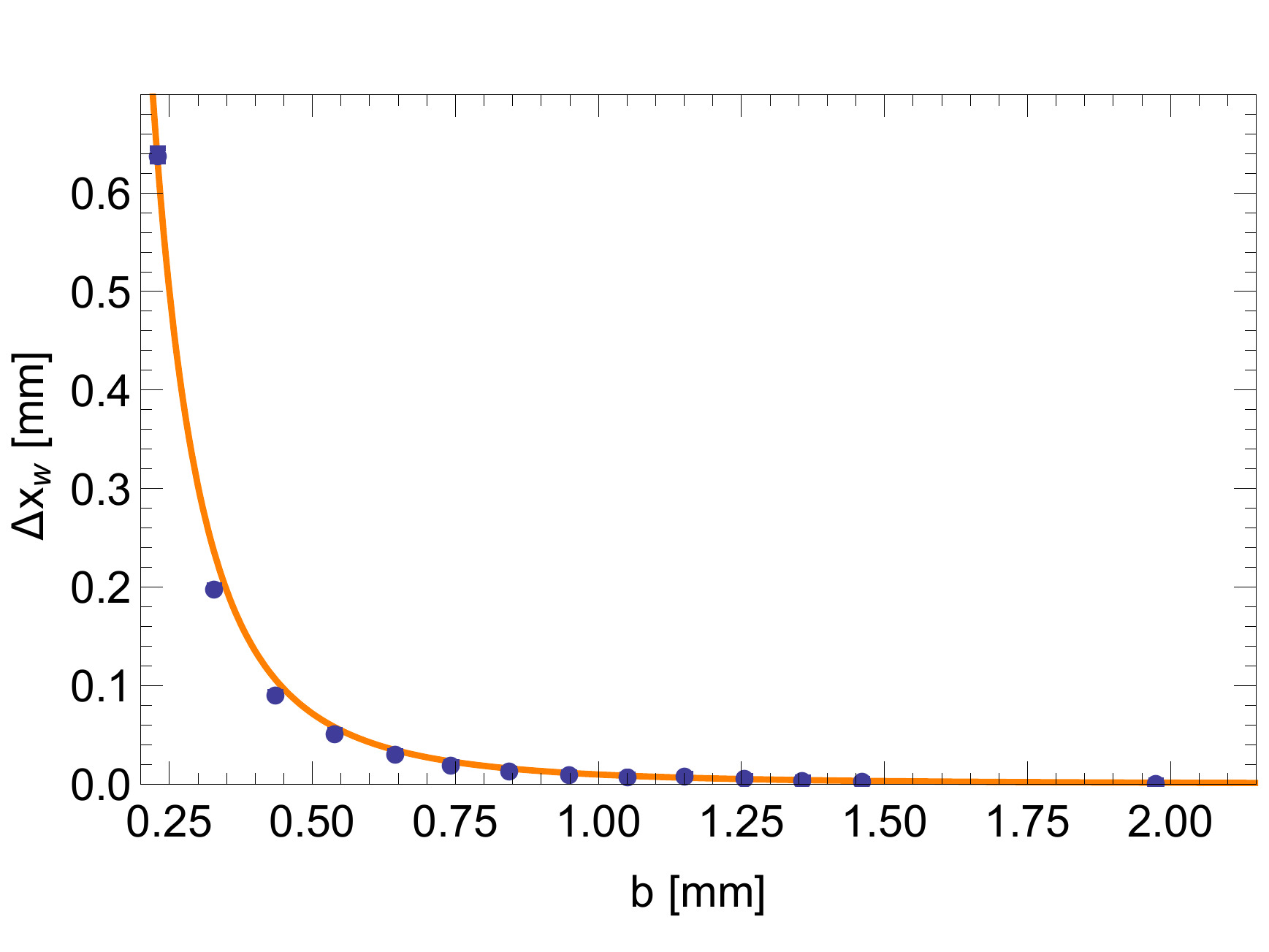}
    \includegraphics[draft=false, width=.46\textwidth]{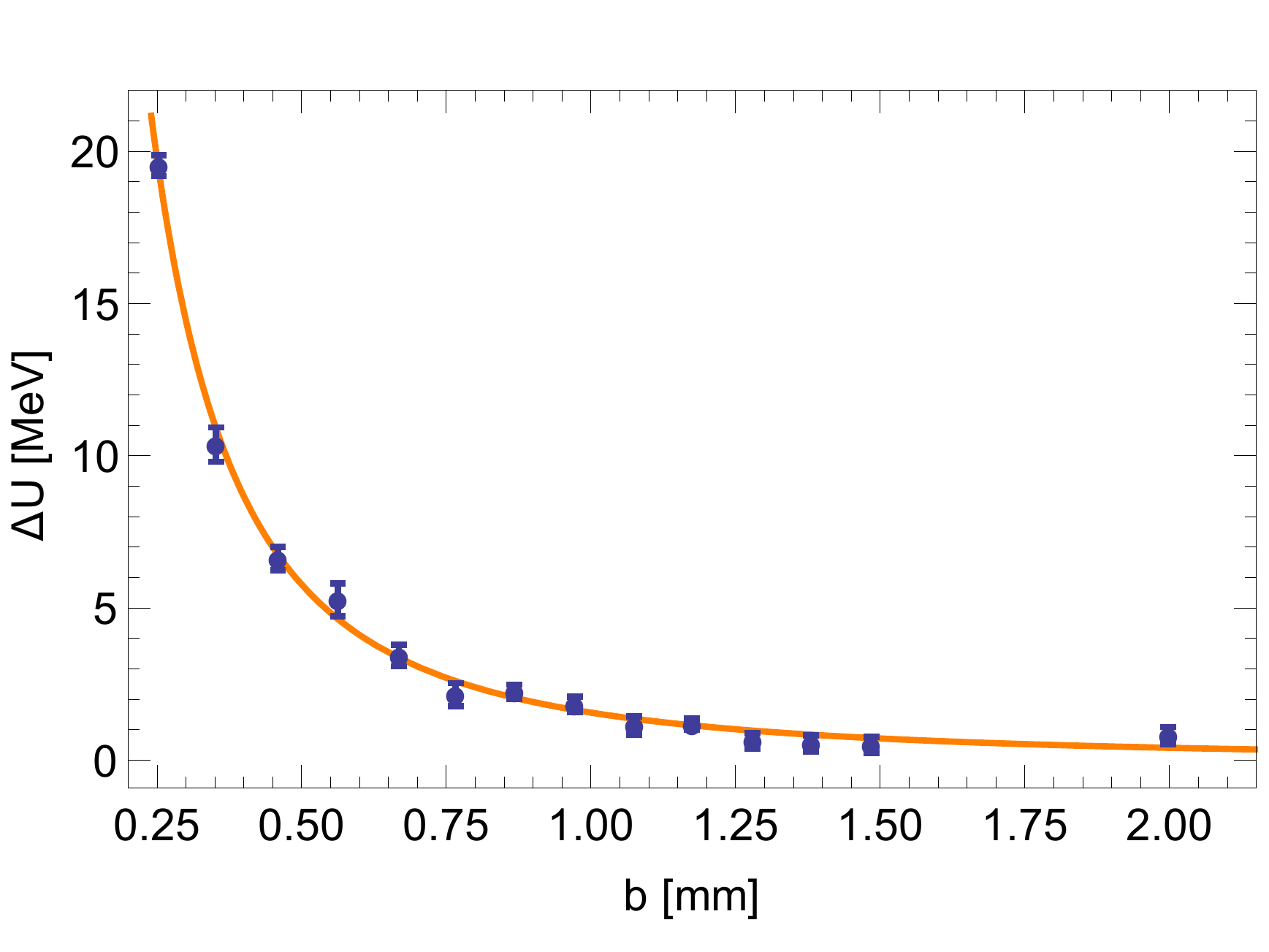}
    \caption{Single north jaw measurements of transverse (top) and longitudinal (bottom) wakefield kicks. The symbols give the data points, with their $b$ values shifted by $-161$~$\mu$m (top) and $-138$~$\mu$m (bottom);  the curves give the analytical theory.}\label{north_kick_fi}
    \end{figure}

\begin{figure}[!htb]
    \centering
    \includegraphics[draft=false, width=.45\textwidth]{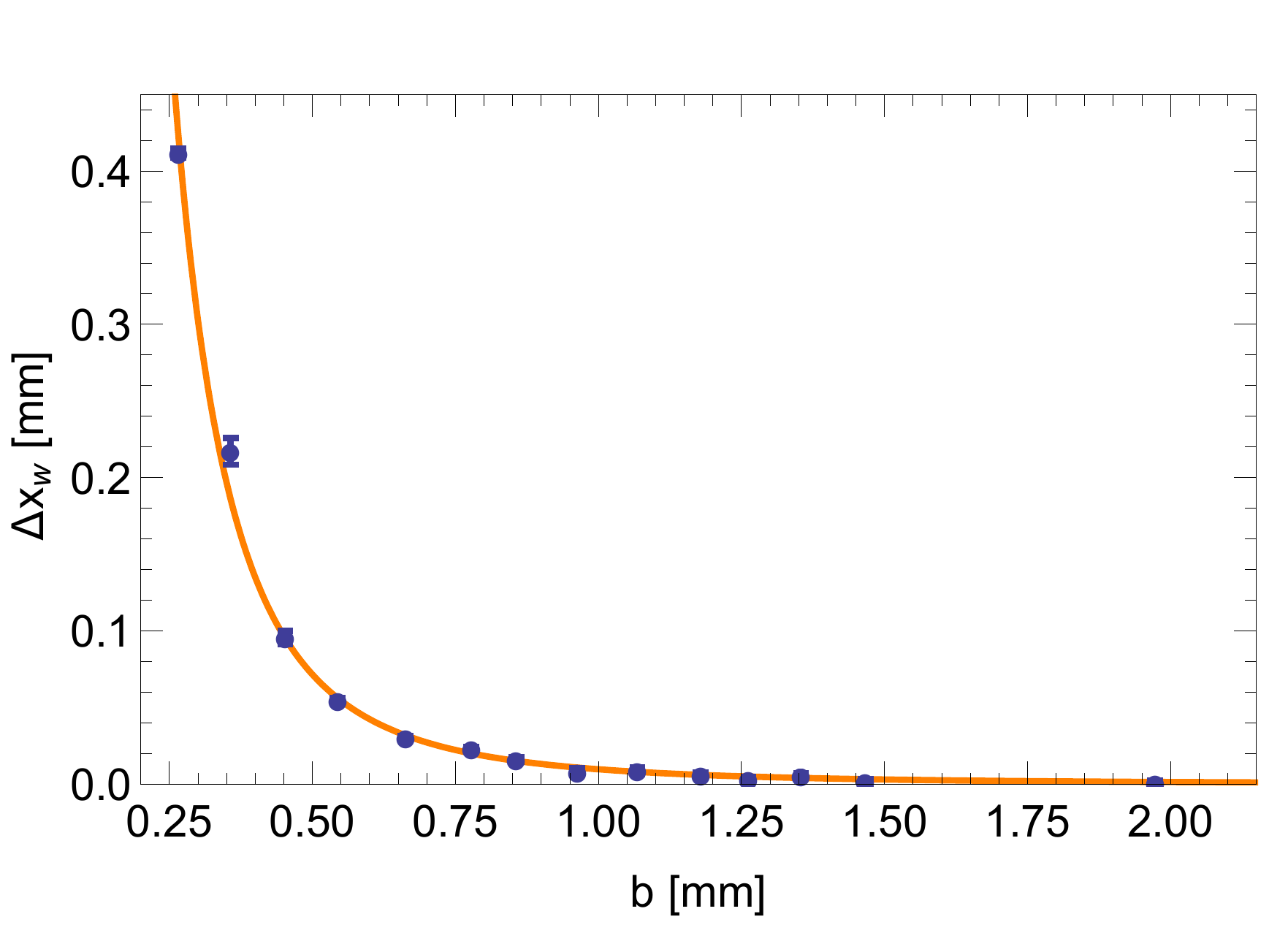}
    \includegraphics[draft=false, width=.46\textwidth]{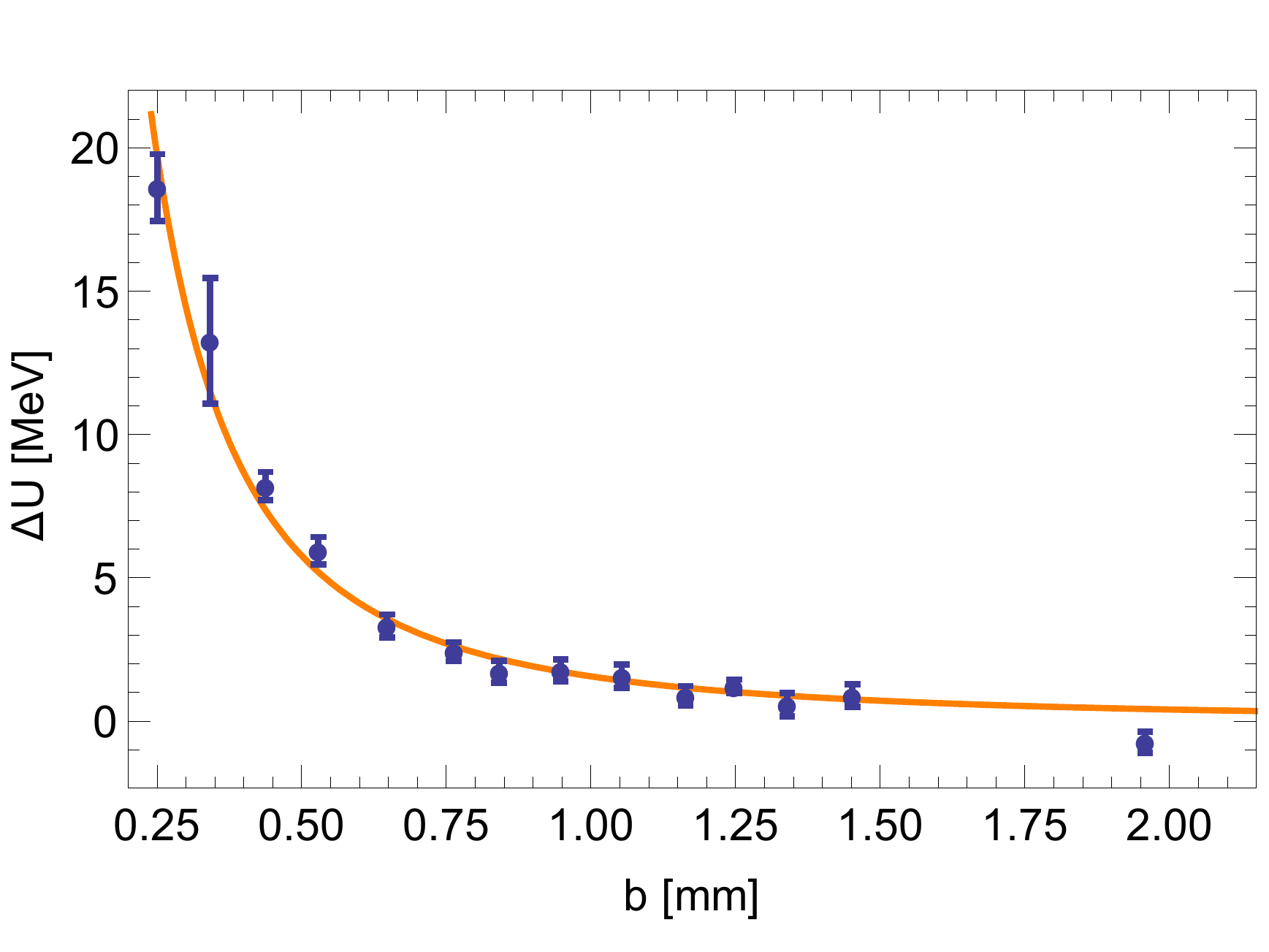}
    \caption{Single south jaw measurements of transverse (top) and longitudinal (bottom) wakefield kicks. The symbols give the data points, with their $b$ values shifted by $-58$~$\mu$m (top) and $-73$~$\mu$m (bottom); the curves give the analytical theory.}\label{south_kick_fi}
    \end{figure}

The measurements were repeated with the south jaw, and the results are shown in Fig.~\ref{south_kick_fi}. We see that, in the energy measurements of the first two points, the standard deviation of error is quite large; this is partly because the number of accepted scans for these points is small, only 10 and 8, respectively. From the figure we see that the fits to the data again are good. This time the fitted shifts are $-58$~$\mu$m (for the transverse kick data) and $-73$~$\mu$m (for the longitudinal data). Again the fitted shifts are in reasonable agreement with each other.

Before leaving this topic, we should mention that we have made two assumptions in the analysis: (1) that the beam passes by the dechirper jaw with little $x$--$z$ tilt, and (2) that the jaw itself is not tilted.
The beam is likely not tilted by much, but as to the jaw tilt, we don't have a good estimate of it. We can just say that the good agreement with the data, and the reasonably good agreement of the fitted shifts for both data sets, are signs that the theory is good and so is our understanding of the dechiper jaw orientation.  The same holds true for the double jaw measurements. Expected deviations are below $100$~$\mu$m.

\section{Conclusions}
\label{sec:conc}

The RadiaBeam/SLAC dechirper
has been installed 
at the Linac Coherent Light Source (LCLS).
Rather than its use as a dechirper, it is the strong transverse kick, induced when the beam passes by off axis, that has received the most interest. This way of using the dechirper has already been employed to {\it e.g.} facilitate the production of intense, two-color photon beams for users~\cite{Alberto}, and is currently being used to develop a diagnostic for
attosecond bunch length measurements~\cite{PhysRevSTAB.18.104402}.  
Thus, for the LCLS, measurements of the strength of the transverse wakefields in this device is an important task that increases our understanding of the dechirper and facilitates our using it in novel ways.
In this report, we present the first systematic measurements of the transverse wakefields in a dechirper, when
%that has been installed in a functioning X-ray FEL, and are 
excited by a bunch  
with the short pulse duration and high energy found in an X-ray FEL.

We have presented measurements of wake kick as function of beam offset for a bunch passing between the two plates of the horizontal dechirper module and compared with analytical calculations. For two example half-gap settings, we found good agreement between theory and measurement, once the aperture (in the calculations) was adjusted by $-11$\% and $-6$\%, respectively. 

We repeated the measurement for the case of a beam passing by each of the jaws of the horizontal dechirper module separately (with the other jaw parked far from the beam). We found good agreement between theory and measurement when we allowed for an overall offset in distance of the bunch from the jaw, $\Delta b$. The fitted offsets were $\Delta b=-161$ ($-58$)~$\mu$m in the north (south) dechirper module jaw. By simultaneously measuring the longitudinal wake effect and performing the corresponding fitting, we found that $\Delta b=-138$ ($-73$)~$\mu$m in the north (south) jaw. We consider this good agreement in offsets between the transverse and longitudinal results.

We have seen good agreement between the measured and calculated wake kicks, after a parameter was slightly adjusted in the calculations: the half-gap $a$ in the case of two-plates and the beam shift in offset $\Delta b$ in the case of a single plate. 
Besides the possibility of inaccuracy in the theory, these discrepancies can be due to measurement error in {\it e.g.} plate offset and plate tilt.
 Nevertheless, the agreement we do see---between measurement and theory---indicates that the theory and our understanding of the dechirper jaw orientation (during the measurement) 
 were both quite good. 

In Appendix~A we derived analytical formulas for the average longitudinal and transverse wake kicks excited by a short, uniform bunch passing through two-plates (or by a single plate) of a dechirper. In analyzing the first, two-plate measurement example, we demonstrated that these formulas agree almost perfectly with the more accurate, convolution method applied to an LCLS-type bunch.

\section*{Acknowledgments}

We thank RadiaBeam for building for us the RadiaBeam/SLAC dechirper, and for their support in installing and commissioning it at the LCLS. We also thank the SLAC operators for helping us make these experiments a success.
Work supported in part by the U.S. Department of Energy, Office of Science, Office of Basic Energy Sciences, under Contract No. DE-AC02-76SF00515  and DE-SC0009550.

\begin{appendix}

%Appendix A

\section{Average Wake for Uniform Bunch Distribution, Analytical Model}

Analytical formulas were developed for the longitudinal and transverse point charge wakes, for the cases of beam between two dechirper plates~\cite{PhysRevAccelBeams.19.084401} and near a single plate~\cite{SLAC-Pub-16881}. These formulas were shown to agree well with numerical simulations. 
For a line-charge or pencil beam (one with small transverse extent) with a uniform longitudinal distribution, it turns out that the average longitudinal and transverse wakefield kicks also are given by simple analytical expressions. We derive these expressions here.

\subsection*{Transverse}

The bunch wake is given by the convolution of the point charge wake and the bunch distribution.
If the longitudinal bunch distribution is uniform of length $\ell$, $\lambda=H(s)H(\ell-s)/\ell$, where $H(s)=1$ (0), for $s>0$ ($<0$), the kick factor---the average of the {\it transverse bunch wake}---is given by
\begin{equation}
\varkappa_x=\frac{1}{\ell}\int_0^\ell w_x(s)\left(1-\frac{s}{\ell}\right)\,ds=\left(\frac{Z_0c}{8\pi}\right)As_{0x}f_x\left(\frac{\ell}{s_{0x}}\right)\ ,\label{kappay_eq}
\end{equation}
where we used $w_x(s)$ as given in Eq.~\ref{wy_eq}; here
\begin{equation}
f_x(\zeta)=1-\frac{12}{\zeta}+\frac{120}{\zeta^2}-8e^{-\sqrt{\zeta}}\left(\frac{1}{\zeta^{1/2}}+\frac{6}{\zeta}+\frac{15}{\zeta^{3/2}}+\frac{15}{\zeta^2}\right)\ .
\end{equation}
%We measure the average (over the bunch) offset at a downstream BPM, given by $\Delta x_w=eQLL_{BPM}\varkappa_x /E$, with $e$ the charge of the electron, $Q$ the charge of the bunch, $L$ the length of the dechirper plates, and $E$ the bunch energy.

\subsection*{Longitudinal}

The longitudinal point charge wake, for a beam between two dechirper plates separated by distance $2a$, or offset a distance $b$ from a single plate, is given in the form~\cite{PhysRevAccelBeams.19.084401}
\begin{equation}
w_z(s)=\left(\frac{Z_0c}{4\pi}\right)Be^{-\sqrt{s/s_{0l}}}\ ,
\end{equation}
with constants $B$ and the longitudinal scale factor
$s_{0l}$. 
In the case of a beam between two plates, the constants are given by 
\begin{eqnarray}
B_{d}&=&\frac{\pi ^2}{ 4a^2}\sec^2\left(\frac{\pi x}{2a}\right)\ ,\nonumber\\
 s_{0d}&=&4s_{0r}\left[1+\frac{1}{3}\cos^2(\frac{\pi x}{2a})+(\frac{\pi x}{2a})\tan(\frac{\pi x}{2a})\right]^{-2}\ .
\end{eqnarray}
In the case of a beam near a single plate, the constants are~\cite{SLAC-Pub-16881}  
\begin{equation}
B_{s}=\frac{1}{b^2}\ ,\quad\quad s_{0s}=\frac{2b^2t}{\pi \alpha^2 p^2}\ .
\end{equation}

For a uniform bunch distribution, the loss factor---the average of the {\it longitudinal bunch wake}---is given by
\begin{equation}
\varkappa=\frac{1}{\ell}\int_0^\ell w_z(s)\left(1-\frac{s}{\ell}\right)\,ds=\left(\frac{Z_0c}{4\pi}\right)Bf_z\left(\frac{\ell}{s_{0l}}\right)\ ,\label{kappa_eq}
\end{equation}
with
\begin{equation}
f_z(\zeta)=\frac{2}{\zeta}\left(1-\frac{6}{\zeta}\right)+e^{-\sqrt{\zeta}}\left[\frac{4}{\zeta}\left(1+\frac{3}{\zeta}\right)+\frac{12}{\zeta^{3/2}}\right]\ .
\end{equation}
%We measure the average energy loss of the bunch, which is given by $\Delta U=eQL\varkappa$.

\end{appendix}

%\bibliography{Dechirper_mainb}% Produces the bibliography via BibTeX.

%\end{document}
%\input{Dechirper_maino.bbl}


%\begin{thebibliography}{99} % Use for 10-99 references
%\input{bibitems.tex}
%\end{thebibliography}

\end{document}